\documentclass[aps,prd,twocolumn]{revtex4-1}

\usepackage{graphicx}

\newcommand{\be}{\begin{equation}}
\newcommand{\ee}{\end{equation}}
\newcommand{\ba}{\begin{eqnarray}}
\newcommand{\ea}{\end{eqnarray}}

\begin{document}

\title{The existence of a two-solar mass neutron star \\
constrains the gravitational constant $G_N$ at strong field}
\date{\today}

%% Notice placement of commas and superscripts and use of &
%% in the author list

\author{Antonio Dobado and Felipe J. Llanes-Estrada}
\altaffiliation[On leave at: ]{Theor. Phys. Dept., Technische Universit\"at Muenchen, 85747 Garching, Germany.}
\affiliation{Departamento de F\'{\i}sica Te\'orica I, Univ. Complutense de Madrid, 28040 Madrid, Spain.} 
%\author{}
%\affiliation{Departamento de F\'{\i}sica Te\'orica I, Univ. Complutense de Madrid, 28040 Madrid, Spain.} 

\author{Jose Antonio Oller}
\affiliation{Departamento de F\'{\i}sica, Universidad de Murcia, 30071 Murcia, Spain.}

\begin{abstract}
In General Relativity there is a maximum mass allowed for neutron stars that, if exceeded, entails their collapse into a black hole. Its precise value depends on details of the nuclear matter equation of state about which we are much more certain  thanks to recent progress in low-energy effective theories.
The discovery of a two-solar mass neutron star, near that maximum mass, when analyzed with modern equations of state, implies that Newton's gravitational constant in the star cannot exceed its value on Earth by more than 8\% at 95\% confidence level. This is a remarkable leap of ten orders of magnitude in the gravitational field intensity at which the constant has been constrained.
\end{abstract}
\maketitle
%%%%%%%%%%%%%%%%%%%%%%%%%%%%%%%%%%%%%%%%%%%%%%%%%%%%%%%%%%%%%%%%%%%%%%%%%%%

The gravitational attraction force between two bodies of mass $m_1$ and $m_2$ at distance $r$ is, if not much precision be required, given by the renowned law of Newton
\be
F = G_N \frac{m_1 m_2}{r^2}\ .
\ee
The Newtonian constant $G_N$, first measured by Cavendish,
also features in the more precise field equations of Einstein's General Relativity
\be \label{Einstein}
R_{\mu \nu} - \frac{1}{2}Rg_{\mu\nu} = 8\pi G_N T_{\mu\nu}
\ee
that need to be used under intense gravitational fields with Einstein's curvature tensor $R_{\mu \nu} - \frac{1}{2}Rg_{\mu\nu}$ and matter source $T_{\mu\nu}$
or otherwise if high precision is expected.
This constant has been carefully measured on numerous occasions~\cite{Gillies87,Mohr05} and is currently taken to be $6.6738(8)N(m/kg)^2$.

\begin{figure}[h]
\includegraphics[width=8cm]{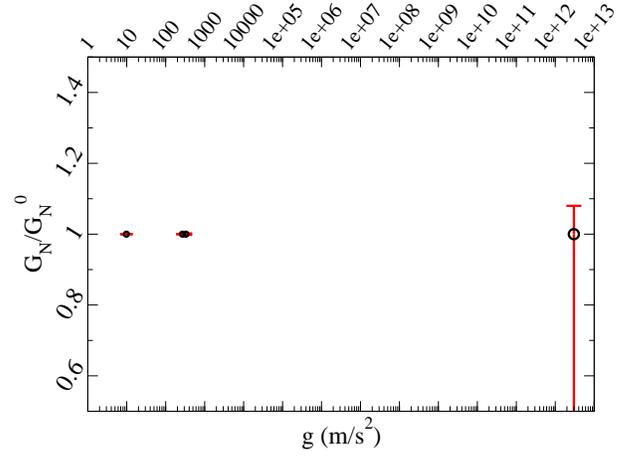}
\caption{The gravitational constant remains (so far) a constant.
Newton-Cavendish constant normalized by its accepted value $6.6738(8)N(m/kg)^2$.  Left point: laboratory on Earth. Middle: orbital determinations of binary pulsars. Right: existence of a neutron star with mass 1.97(4) solar masses. At the intense gravitational field in such neutron star, the gravitational constant cannot exceed $8\%$ of its value on Earth at $95\%$ confidence level.\label{fig:Cavendish}}
\end{figure}

The study of the orbital evolution of binary pulsars~\cite{Kramer:2006nb} has also allowed to establish the validity of General Relativity (and indirectly, provided a measurement of $G_N$) under stronger gravity conditions. While on the Earth's surface $g\simeq 9.8\ m/s^2$, for the binary pulsar J0737-3039 the relevant acceleration is $g\simeq 270\ m/s^2$, and for PSR B1913+16, $330\ m/s^2$. The assessment of $G_N$ in those systems has a respective precision of $0.05\%$ and $0.2\%$ (from measurements of corrections to Kepler's law, particularly a parameter called $s$ that is proportional to $G_N^{-1/3}$).

 We here point out that the new discovery~\cite{demorest2011} of a neutron star with a mass equal to 1.97(4) solar masses (a convenient unit weighing about $2\times 10^{30}\ kg$), confirming previous claims of neutron stars in this mass range~\cite{Nice:2005fi} is so close to the maximum mass that such an object can have~\cite{Lattimer:2000nx}, by nuclear physics considerations, that it constrains the value that the gravitational constant can take in its interior. The reason is that there is an equilibrium between gravitational attraction and interneutron repulsion at short distances, that cannot be maintained somewhat above two solar masses, and heavier objects collapse into black holes. Thus an increase of the gravitational constant $G_N$ that diminishes the maximum mass attainable, is excluded by the discovery of these superheavy neutron stars (most neutron stars known to date had masses near 1.4 times that of the Sun).

The situation is described in figure~\ref{fig:Cavendish}.
From left to right, we show the laboratory value  (precision $1\times 10^{-4}$); the astronomical value as inferred from double pulsars J0737-3039 and Hulse-Taylor PSR B 1913+16, at acceleration near $300\ m/s^2$ (respective precisions $5\times 10^{-4}$ and $2\times 10^{-3}$), and our constraint from the existence of a neutron star with mass equal to 1.97(4) solar masses ($G_N$ cannot exceed its earthly value by more than $8\%$ at $95\%$ confidence level).

\begin{figure}[h]
\includegraphics[width=8cm]{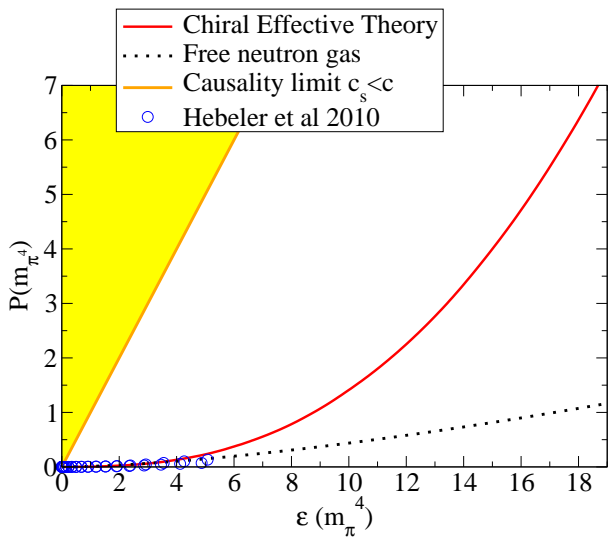}
\caption{The equation of state (pressure as function of density) for pure neutron matter (solid red line), compared with the free neutron gas (dashed black line) is much stiffer because of repulsive interactions (nuclear matter is barely compressible) and an independent low-density effective theory equation of state~\cite{Hebeler:2010jx}. Also shown is the causality limit, given by the condition that sound propagates slower than light, $c_s^2 = \partial P/\partial \rho <c$.  \label{fig:Pofrho}}
\end{figure}

To obtain the bound we employ the equation of hydrostatic equilibrium of Tolman-Oppenheimer-Volkoff~\cite{Tolman:1939jz,Oppenheimer:1939ne}, consequence of Eq.~(\ref{Einstein}). This equation governs the variation of the pressure $P$ inside a spherically symmetric, static star at radial distance $r$ from its center, 
inside which a mass $M(r)$, due to the
mass-energy density $\varepsilon(r)$, has accumulated
\be \label{TOV}
\frac{dP}{dr} = - \frac{G_N}{r^2}
\frac{(\varepsilon(r)+P(r))(M(r)+4\pi r^3P(r))}{1-\frac{2G_NM(r)}{r}} \ .
\ee
This equation can be integrated numerically from the inside of the star ($r=0$) to the outside by a standard Runge-Kutta computer algorithm. The initial condition is supplied as a value of the pressure in the star's center, and the equation is considered solved at the distance $R$ where the pressure drops to zero. $R$ is then interpreted as the star radius.
The mass function $M(r)$ and total star's mass $M(R)$ are obtained by adopting an equation of state that relates the total energy density $\varepsilon(r)$ to the pressure $P(r)$. Here is where steady progress in nuclear theory allows us to have reasonable confidence in the equation of state input, shown as the solid red-line in figure~\ref{fig:Pofrho}.

The plot also shows, for comparison, the equation of state for a pure neutron Fermi gas~\cite{Zeldovich}. This is much less ``stiff'' (lower pressure at given energy density) since it does not include the interneutron interaction, that is repulsive. Further, we show an independent computation of the equation of state at low energy densities that is in good agreement~\cite{Hebeler:2010jx} with our own computation. A slight discrepancy can be adscribed to our employing pure neutron matter for simplicity while those authors are including a small amount of protons in dynamical $\beta$-equilibrium with the neutrons. 
Our improvements concern the use of a recently developed chiral effective field theory (EFT) for nuclear matter~\cite{Lacour:2009ej,Oller:2009zt,Lacour:2010ci,Meissner:2001gz} based on a new power counting. The
latter is able to single out the set of Feynman diagrams, including infinite strings of them, that are required to calculate order by order in the
chiral expansion for (asymmetric) nuclear matter. This is a novelty
since previous calculations employ standard many-body methods not based in the new paradigm of  EFT. The chiral power counting~\cite{Lacour:2009ej} considers multi-nucleon forces both from pion exchanges as well as from short ranges contributions. It also takes into account the important infrared enhancement affecting nucleon propagators in the multi-nucleon reducible loops. This allows to control the size of many-body (three-body, four-body, etc.) forces and decide, systematically and to the precision desired, what medium effects are to be kept at each stage of the calculation.

\begin{figure}[h]
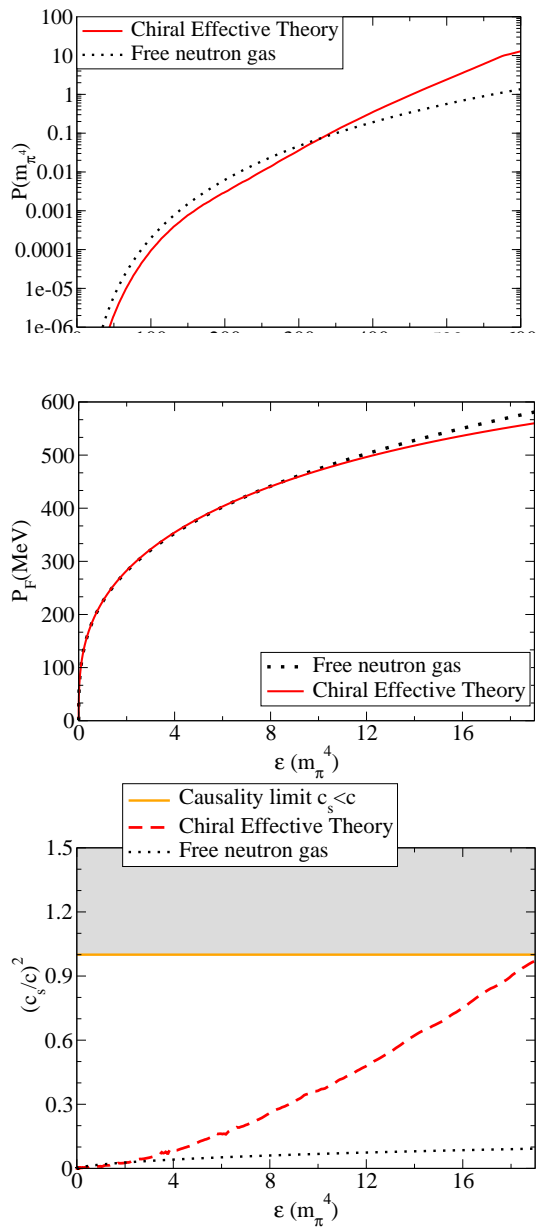

\includegraphics[width=7cm]{fig3a.eps}\vspace{0.2cm}
\includegraphics[width=7cm]{fig3b.eps}
\includegraphics[width=7cm]{fig3c.eps}
\caption{
Equation of state for pure neutron matter in effective theory (red solid line) versus the free neutron gas (black dashed line). Top: Pressure against Fermi momentum. Middle pannel: density as function of the Fermi momentum. Bottom: speed of sound showing the causality limit where the effective theory description is expected to break down. \label{fig:moreEqState}}
\end{figure}

Further details on the equation of state that we employ are documented in Figure~\ref{fig:moreEqState}, that presents the pressure and energy densities as functions of the Fermi momentum for the neutron gas, as well as the speed of sound $\sqrt{\partial P/\partial \rho}$.
It should be noted that, while obtained with sophisticated modern effective field theory treatments, the equation of state is in broad agreement with vintage nuclear theory treatments based on phenomenological potentials~\cite{Akmal:1998cf}.

\begin{figure}[h]
\includegraphics[width=8cm]{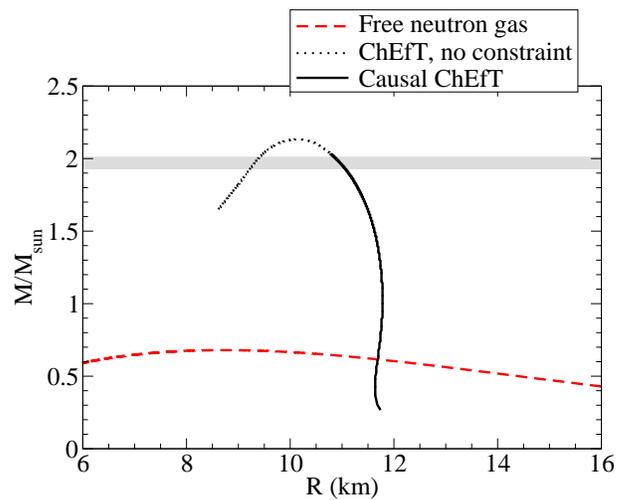}
\caption{
Mass-radius plot for the neutron star solutions of the Tolman-Oppenheimer-Volkoff equation of hydrostatic equilibrium. As is well known, the free neutron gas equation of state cannot reach masses beyond about 0.6 solar masses. However the interacting equation of state, being more repulsive, supports stars slightly above 2 solar masses against gravitational collapse, in agreement with the observation of a star with 1.97(4) solar masses.
\label{fig:MR}}
\end{figure}

\begin{figure}[h]
\includegraphics[width=8cm]{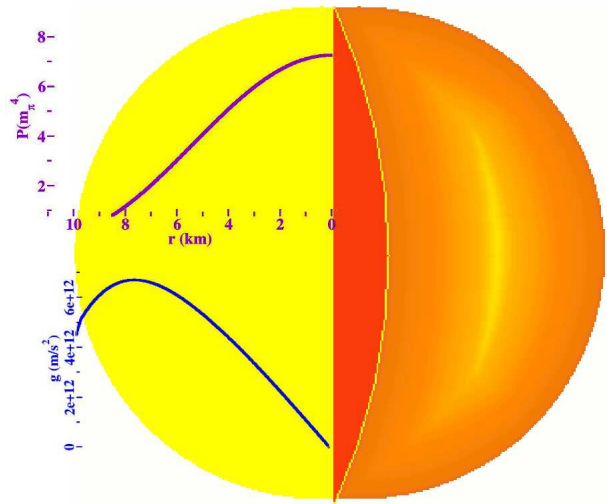}
\caption{For a neutron star with mass near the maximum allowed by hydrostatic equilibrium, we show the pressure profile and the acceleration of gravity (note its non-Newtonian behaviour due to the relativistic pressure term).  
\label{fig:profile}}
\end{figure}

Once the equation of state has been fixed, and the integration of the Tolman-Oppenheimer-Volkoff equilibrium equation~(\ref{TOV}) has proceeded, one obtains the standard mass-radius plot in figure~\ref{fig:MR}.
As is known since the early work of Oppenheimer and Volkoff, a pure neutron gas supports no star with mass above 0.6-0.7 solar masses, providing a check of our computer programme (dashed red line). The full calculation including interactions can elevate the maximum mass above 2 solar masses. We discontinue the solid black line at a point where the effective theory breaks down as manifested by reaching the causality limit \footnote{An additional higher order computation of the sound velocity within the
EFT~\cite{Lacour:2009ej} is planned since the excess over the casuality limit from the thermodynamical formula $c_s^2=\partial P/\partial \rho$ is only at the $10\%$ level. This has no impact in our current results.} $c_s=c$
 (the Fermi momentum at that point, about $600\ MeV$, is also quite high). Stars above 2.2-2.3 solar masses are not supported.
For a star with mass about two solar masses, near the maximum possible, we give in figure~\ref{fig:profile} the profiles of pressure and intensity of gravity from the center of the star. The acceleration of gravity does not grow uniformly from the center to the edge as in Newtonian mechanics (as seen easily from Gauss's law) due to the pressure contribution in the relativistic expression for the potential $\Phi$ (given in geometrodynamic units $c=G_N=1$)
\be
g= \frac{d\Phi}{dr} = \frac{M(r)+4\pi r^3 P(r)}{r(r-2M(r))} \ .
\ee
The order of magnitude of $g$ in the full calculation can be understood from simple Newtonian considerations as $G_N M(R)/R^2$ at the star's surface, and if corrected by the relativistic denominator $(1-2G_NM(R)/c^2)^{-1}\simeq 1.4$ one obtains about $g\simeq 2\times 10^{12}\ m/s^2$.
\footnote{Since the pulsar's period is measured to be about 3.15 milliseconds, the maximum centripetal acceleration a the Equator is two orders of magnitude smaller than gravity, and we therefore neglect the (naturally) very small oblateness of the star.}

Thus, any information gained on $G_N$ extends our knowledge several orders of magnitude in field intensity. \\
This can be of use to constrain modified theories of gravity, motivated by string theory and by cosmology~\cite{delaCruzDombriz:2009xk}, that suggest that for largely different values of the curvature $R$ (or acceleration field $g$), gravity separates from its Einstenian formulation.

We complete our analysis by returning to Eq.~(\ref{TOV}) and varying $G_N$. Since reducing it simply delays gravitational collapse and eventually allows for arbitrarily heavy stars, no constraint is put in smaller-than-physical $G_N$ values, as reflected in figure~\ref{fig:Cavendish}. However, increasing $G_N$ rapidly reduces the maximum possible mass of the neutron star.

\begin{figure}[h]
\includegraphics[width=8cm]{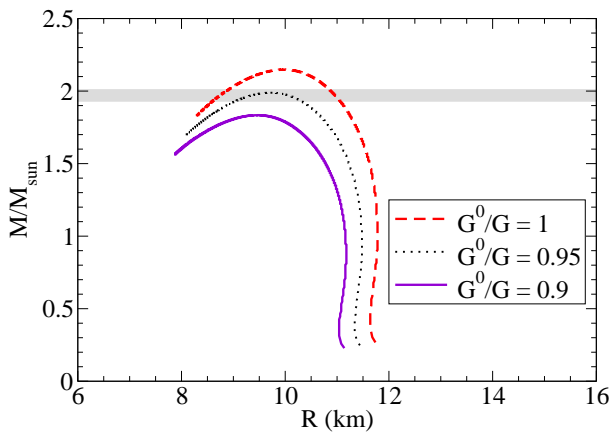}
\caption{Mass-radius plot as in figure \ref{fig:MR} but varying the gravitational Newton-Cavendish constant. As this grows (or, as shown, the ratio of the constant in Earth to that constant at high field decreases) the star becomes more prone to gravitational collapse and thus the maximum reachable mass drops below the requisite two solar masses. Thus, increases of $G_N$ are now constrained.
\label{fig:MRfinal}
}
\end{figure}

To control the systematic uncertainty at very high energy densities, where other phenomena might arise (activation of the strangeness degree of freedom, transition to a different phase of nuclear matter not accessible from the nucleon effective theory, etc.), and since we are interested in imposing \emph{an upper bound} on $G_N$, we substitute our equation of state by the stiff-most allowed by causality\footnote{Introducing additional, possibly exotic, degrees of freedom cannot stiffen the equation of state beyond causality.}, such that $c_s=c$, yielding $P=c^2 (\rho-\rho_{\rm max}) + P_{\rm max}$,
above a maximum Fermi momentum of either $k_F=600$ or $450\ MeV$. Shown in figure~\ref{fig:MRfinal} is the mass/radius plot adopting the first value, allowing for the Cavendish constant to vary. From this calculation we derive the bound on an 8\%
variation of $G_N$.\\
Should one adopt the second value due to putative errors that we may have not identified in our equation of state at higher energy density, the constrain
on $G_N$ is somewhat relaxed, but remains meaningful, excluding a variation of 19\% at the $2\sigma$ level. \\
On the other end, we neglect corrections due to the neutron star skin containing several atomic sheets~\cite{LlanesEstrada:2009na} as well as ordinary nuclear (not neutron) matter, for it is known that its total contribution to the star's mass is rather small~\cite{crust1995,Hebeler:2010jx}.

In conclusion, we believe that we have made a relevant contribution in employing the new two-solar mass neutron star to constrain the gravitational constant $G_N$ given hard information on the nuclear matter equation of state, as opposed to attempting to constrain the equation of state and possibly exotic forms of nuclear matter about which abundant literature exists~\cite{Chowdhury:2011cf,Li:2011vd,Wen:2011rb,Ozel:2010fw,Lattimer:2010uk}.\\
Other authors~\cite{Hebeler:2010jx} have already pointed out that nuclear physics is precise enough to constrain the radius of the star given its mass. The discovery of higher-mass neutron stars allows us to establish limits on allowed variations of gravity itself in a hitherto unexplored regime.\\
Not long ago, a two-solar mass neutron star was thought unlikely~\cite{Bethe:1995hv}. Although we now know that nuclear physics can accommodate it, the margin is narrow and indeed allows to constrain gravity.\\
Future work may include an examination of modified theories of gravity (for the case of scalar modifications of the action in $f(R)$ theories the Tolman-Oppenheimer-Volkoff equations are already available in the literature~\cite{Kainulainen:2006wz}).

\begin{acknowledgments}
FJLE thanks a Caja Madrid fellowship for advanced study and the hospitality of the theory group at TU-Munich and the Exzellenzcluster „Origin and Structure of the Universe“.  This work has been supported by grants
 227431-HadronPhysics2 (EU), Consolider-CSD2007-00042, AIC10-D-000582,       FPA2008-00592, FIS2008-01323, FPA2010-17806, 11871/PI/09 (Fundaci\'on S\'eneca, Murcia) and UCM-BSCH GR58/08 910309 (Spain).
\end{acknowledgments}

\bibliography{bindG}
\end{document}